\documentclass[a4paper,aps,pra,twocolumn,preprintnumbers,amsmath,amssymb]{
revtex4}

\usepackage{amsmath}
\usepackage{verbatim}
\usepackage{graphicx}
\usepackage{color}

 \newcommand{\ket}[1]{\left|#1\right\rangle} 
 \newcommand{\bra}[1]{\left\langle#1\right|} 

\begin{document}


\title{Chains with loops - synthetic magnetic fluxes and topological order in one-dimensional spin systems}

\author{ Tobias Gra\ss$^1$, Christine Muschik$^{1,2,3}$, Alessio Celi$^1$,
Ravindra Chhajlany$^{1,4}$, Maciej Lewenstein$^{1,5}$}

\affiliation{$^1$ICFO-Institut de Ci\`encies Fot\`oniques, Av. Carl Friedrich
Gauss 3, 08860 Barcelona, Spain}
\affiliation{$^2$Institute for Quantum Optics and Quantum Information of the
Austrian Academy of Sciences, A-6020 Innsbruck, Austria}
\affiliation{$^3$Institute for Theoretical Physics, University of Innsbruck,
A-6020
Innsbruck, Austria}
\affiliation{$^4$ Faculty of Physics, Adam Mickiewicz University, Umultowska
85, 
61-614 Pozna{\'n}, Poland}
\affiliation{$^5$ICREA-Instituci\'o Catalana de Recerca i Estudis Avan\c cats,
Lluis
Campanys 23,
08010 Barcelona, Spain}

\begin{abstract}
Engineering topological quantum order has become a major field of physics. Many advances have been made by synthesizing gauge fields in cold atomic systems.
Here, we carry over these developments to other platforms which are extremely well suited for quantum engineering, namely trapped ions and nano-trapped atoms.
Since these systems are typically one-dimensional, the action of artificial magnetic fields has so far received little attention. However,
exploiting the long-range nature of interactions, loops with non-vanishing magnetic fluxes become possible even in one-dimensional settings.
This gives rise to intriguing phenomena, such as fractal energy spectra, flat bands with localized edge states, and topological many-body states. 
We elaborate on a simple scheme for generating the required artificial fluxes by periodically driving an XY spin chain. Concrete estimates demonstrating the
experimental feasibility for trapped ions and atoms in waveguides are given.
\end{abstract}

\maketitle

\section{Introduction}
When an electric charge moves on a closed loop in the presence of a magnetic field, its wavefunction picks up a geometric phase according to the magnetic
flux through the contour. This can give rise to intriguing phenomena, such as the famous fractal energy spectrum, known as Hofstadter butterfly \cite{hofstadter}, predicted for particles on a two-dimensional lattice. Apart from its esthetical appeal, the intimate relation to topological order has triggered immense interest in this phenomenon. In recent years, engineering topological quantum systems with cold atoms in artificial gauge fields has become a major field of research \cite{jakschzoller,osterloh,clark,spielmanPRL,spielman-peierls,powellPRL, goldmanTI,trombettoni,hauke-shakes,sengstock12,bloch-hofstadter, ketterle-harper,Celi13,grass-anyons,dalibard,Goldman13}.

\begin{figure}[t]
\centering
\includegraphics[width=0.45\textwidth, angle=0]{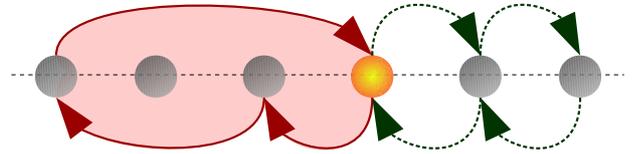} 
\caption{\label{fig0} (Color online)
Loops in a one-dimensional lattice: The green loop (dotted) is trivial, as phases on forward and backward links would always cancel each other. The situation is different for the red loop, where long-range connections are exploited.
}
\end{figure}

However, since the effect of a magnetic field is trivial in one spatial dimension, as any loop encloses zero flux, promising platforms such as trapped ions 
\cite{porras04,blattross} or atoms coupled to waveguides  \cite{darrick,kimble14,Vetsch2010,Goban2012} have been excluded from quantum simulations of the Hofstadter model.
On the other hand, an interesting property offered by these systems are long-range spin-spin interactions. From a formal perspective, a spin flip interaction is analog to the hopping of a particle.  Accordingly, a complex-valued spin flip interaction represents the hopping of a charged particle in the presence of a magnetic field. 
While in a short-ranged system, back- and forward hopping would simply cancel the magnetic flux through any loop, the presence of long-range connections changes the situation, allowing for non-trivial loops as illustrated in Fig. \ref{fig0}. In this article, we elaborate on this idea, and design a spin chain model with non-vanishing magnetic fluxes. We show that it exhibits a Hofstadter-like fractal energy spectrum, and reveal topological phases by calculating edge states and Chern numbers. For the engineering of the model with trapped ions or atoms coupled to waveguides, we resort to simple periodic driving techniques, which allow for simultaneous control over strengthes and complex phases of individual spin-spin interactions.

\begin{figure}[t]
\centering
\includegraphics[width=0.45\textwidth, angle=0]{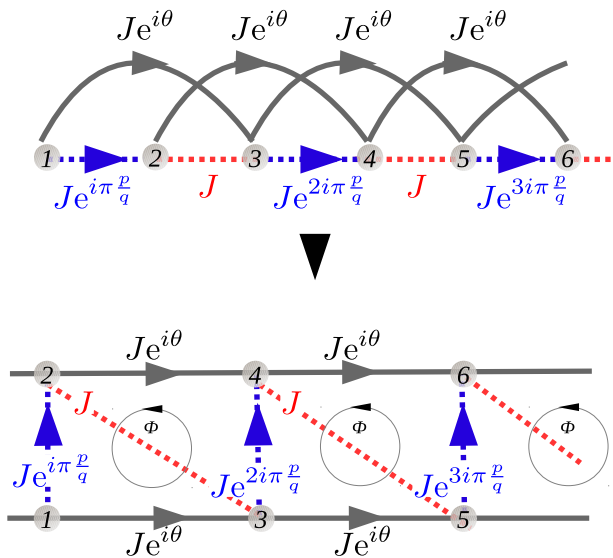} 
\caption{\label{fig1} (Color online)
A spin chain with nearest-neigbor interactions and next-nearest neighbor interactions (above) is mapped onto a triangular ladder (below). The shown setup leads to constant fluxes $\Phi=\pi \frac{p}{q}$ through square plaquettes.
}
\end{figure}

The approach to topological order envisaged in this article is based on three observations: First, we resort to the formal analogy between the hopping of a particle and a spin-flip interaction. Second, we notice that by assigning complex coupling strengthes to the spin-spin interactions they can mimic the hopping of charged particles within a magnetic field. Finally, we exploit the long-range character of spin-spin interactions in systems of trapped ions or atoms coupled to nanofibers. This is the crucial ingredient which allows us to design spin chains with loops of non-zero ``magnetic'' fluxes. In our picture, a closed loop is given by a sequence of spin flips which returns to the initial configuration. The flux is given by a summation over the complex phases of each interaction. While the general idea of this scheme holds for any long-range interaction, an appealing and geometrically simple interpretation is possible for spin chains with only nearest-neighbor (NN) and next-to-nearest-neighbor (NNN) interactions: As illustrated in Fig. \ref{fig1}, such chain can be mapped onto a triangular ladder with NN interactions, indicating an intriguing analogy between dimensionality and the range of interactions. 

An interesting feature of our mapping is the possibility of controlling the number of particles via the spin polarization. A sufficiently strong magnetic field which polarizes all but one spins leads to the realization of single-particle physics. In the presence of more than one spin flip, the spin chain maps onto a system of strongly interacting hard-core bosons.
In both the single-particle and the many-body configurations, it is natural to ask for topological phenomena, if artificial magnetic fluxes are attached to the spin chain. Our respective analysis starts by considering the strongly polarized chain, with only one spin flipped relatively to the others. In this single-particle limit, our spin chain model directly realizes the Hofstadter model on a triangular ladder. Strikingly, and in contrast to a square ladder, the triangular geometry supports the occurrence of a butterfly-like fractal energy spectrum. We also find localized edge states \cite{kraus12,lang12,marra}, and it is possible to define topological invariants taking non-zero values. These findings suggest topological properties similar to the integer quantum Hall effect, but one has to bear in mind the one-dimensional nature of our setup. According to the general classification scheme \cite{kitaev}, this requires protection of topological order by some symmetry. In the proposed setting, however, we have topological order in the absence of any obvious symmetry, just as for a two-dimensional system. This is another hint for an increased dimensionality due to the long-range character of the interactions.
By decreasing the spin polarization, our model captures the physics of strongly interacting bosons. Depending on the precise values for filling factor and artificial magnetic fluxes, the topological properties may persist in the many-body scenario: As indicated by Chern number calculations, the bosonic ground state is topologically equivalent to a many-body state obtained from filling the single-particle bands, similar to the recently found interacting integer quantum Hall phases \cite{senthil2013,furukawa2013,wu2013,davidPRB}.

The implementation of the spin chain envisaged in this paper requires control over strengthes and complex phases of the spin-spin interactions. Flexible systems with generically long-ranged spin-spin interactions are available in trapped ions \cite{porras04,blattross}, or in atoms coupled to photonic waveguides \cite{kimble14}. In order to custom-tailor arbitrary interaction strengthes, we propose a scheme inspired by lattice shaking methods. They were introduced for cold atoms in optical lattices~\cite{Eckardt2005,andre07,lignier07,kierig08,sias08,zenesini09,hemmerich10,struck11,sengstock12,hauke-shakes,struck13,jotzu14,goldman14,baur14,struck14}, and have also been applied to ions in 2D microtraps~\cite{bermudez2011}. In this paper, we develop a driving protocol which is able to simultanously adjust absolute value and complex phase of different couplings. Despite its versability, our protocol is extremely simple, as it consists only of piecewise constant, local energy offsets. It can be implemented with an overhead that is independent of the number of atoms involved.

In summary, the purpose of our article is to introduce a spin model mimicking hopping in the presence of magnetic fields. The long-range character of spin-spin interactions allows us to create loops with non-vanishing flux even in a one-dimensional setting. We study topological properties of this model, in both the single-particle and many-body regime. We find a fractal energy spectrum, localized edge states, and quantum states with non-zero Chern number. Finally, we develop a simple driving protocol which allows for implementing the model in promising platforms such as trapped ions or atoms coupled to waveguides.

\section{Model}
We consider an XY spin chain with possibly long-range, and complex-valued spin-spin interactions $J_{ij}$, and a polarizing transverse field $h$:
\begin{align}
\label{XY}
H = - \sum_{i\neq j} \left(J_{ij} \sigma_i^+
\sigma_j^-
+ {\rm H.c.} \right) + h \sum_i \sigma_i^z,
\end{align}
This Hamiltonian commutes with the spin polarization $S_z=\sum \sigma_i^z$, so we can work in sectors of fixed $S_z$. The last term of the Hamiltonian then reduces to a constant controlling the spin polarization. For sufficiently strong $|h|$, all but one spins are aligned in $z$-direction, that is $S_z=N-2$, where $N$ is the total number of
spins. In this subspace, the spin-flip interaction of the XY model maps onto a
free hopping model of a single particle in a lattice. For $|S_z|<N-2$, that is in the many-body regime, a mapping
onto a free fermion model can be achieved via a Jordan-Wigner transformation 
 if interactions are restricted to nearest neighbors. In the
presence of long-range interactions, the Jordan-Wigner transformation additionally produces 
occupation-dependent tunneling terms. Thus, it is more convenient to interpret the long-range spin chain as a system of hardcore bosons. 

A key ingredient to our model are complex phases in the interaction parameters, $J_{ij}=|J_{ij}| \exp(i \varphi_{ij})$,  mimicking the minimal coupling to a vector potential. Moreover, long-range interactions allow for mapping the chain onto a graph embedded in larger dimension. In this paper, we will assume sufficiently fast decaying interactions, such that the model is restricted to NN and NNN interactions.  As illustrated in Fig. \ref{fig1}, the chain can then be mapped onto a triangular ladder where each site is back- and forward connected to a NNN via the horizontal bounds, as well as to the two NNs via the other bounds. By including, for instance, also third-neighbor interactions, the chain would represent an arrangement of linearly connected tetraeders, that is, a geometrical structure embedded in three dimensions.

We focus on the $J_1-J_2$ model, with $|J_1|=|J_2|=J$, and complex-valued parameters. Both, complex phase and strength of the couplings are adjusted by periodic driving techniques (see Section IV ``Experimental implementation'' and Appendix B). Viewed as a triangular ladder, the elementary plaquettes are given by three spins $i$, $i+1$, and $i+2$ forming a triangle. The magnetic flux associated to the $i$th triangle is $\Phi_i = (-1)^i \left(\varphi_{i+1,i}+\varphi_{i+2,i+1}+\varphi_{i,i+2} \right)$. Through gauge invariance, the fluxes $\Phi_i$ uniquely define the physics. As a
convenient choice, we will set phases on NNN links to a constant $\theta$,
$\varphi_{i+2,i}=\theta$, while every second NN bond is equipped with a
spatially dependent phase: $\varphi_{i+1,i}= \frac{\pi(i+1)p}{2q}
\delta_{(i+1){\rm mod}2,0}$, with $p$ and $q$ two coprime integers.
With this choice, the flux through squares, as shown in Fig. \ref{fig1}, is constant: $\Phi=2\pi(p/2q)$.

\section{Results}
\subsection{Fractal energy spectrum}
\begin{figure}[t]
\centering
\includegraphics[width=0.47\textwidth, angle=0]{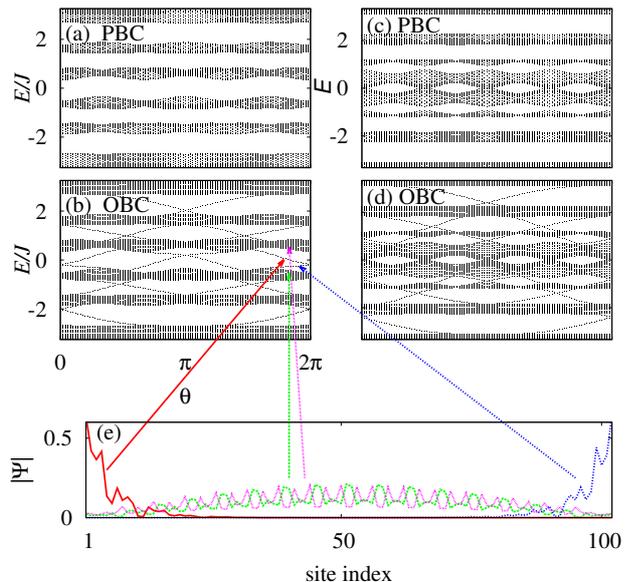}
\caption{\label{fig2} (Color online)
(a--d) Energy spectra (for $S_z=N-2$) as a function of a constant
NNN phase $\varphi_{i,i+2}=\theta$. In (a,b) we consider a system of  $N=102$ spins with a flux
$p/q=1/3$.   In (c,d) we have $N=104$ spins at $p/q=1/4$. In (a,c) we assume periodic boundary conditions, while in (b,d) the boundary is open. In the latter case, edge states fill the gaps. They are localized at the ends of the chain, as seen from the wavefunction amplitude for
selected states plotted in (e).}
\end{figure}

In the single-particle configuration, that is for $S_z=N-2$, the model can be solved by Fourier transformation, see appendix A. As in the Harper/Hofstadter model, the presence of magnetic flux leads to a splitting of the dispersion $E(k)$ into $2q$ bands, restricted to a magnetic Brillouin zone, $k \in [-\pi/2q,\pi/2q]$. Alternatively, the model is solved in real-space by diagonalizing a $N \times N$ matrix. A typical band structure for odd $q$ is shown in Fig. \ref{fig2}(a), with flat bands separated by large gaps. The band structure is slightly different in the case of even $q$,  where two bands touch in Dirac cones at $E=0$, as shown in Fig. \ref{fig2}(c).

\begin{figure}[t]
\centering
\includegraphics[width=0.47\textwidth, angle=0]{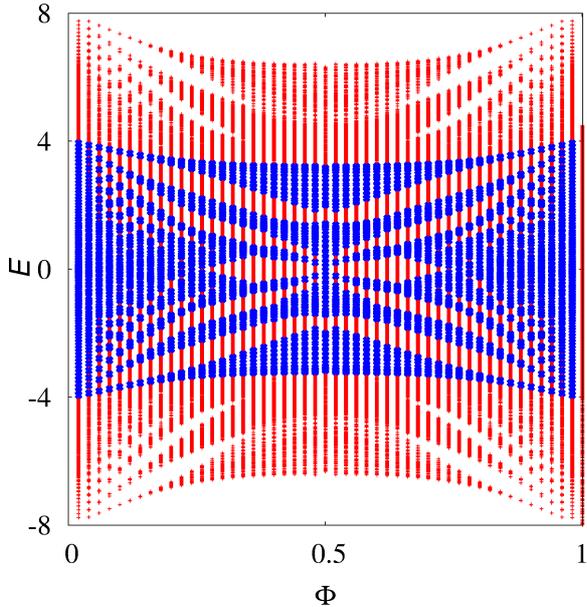}
\caption{\label{fig3} (Color online)
Fractal energy spectrum: the energy levels (in units of $J$) are plotted as a function of the flux $\Phi$ (in units of $2\pi$),  for a single spin flip, $S_z=N-2$ (blue), and two spin flips, $S_z=N-4$ (red). We assume periodic boundary conditions and $\theta=0$ in a system of $N=100$ spins.}
\end{figure}

The occurrence of flat bands in 1D is remarkable. It sensibly depends on the topology of the chain, and most of the gaps disappear if, for instance, the diagonal couplings in Fig. \ref{fig1} (red lines) are set to zero, turning the triangular ladder into a square ladder. Flatness of the bands leads to a fractal figure, similar to the Hofstadter butterfly \cite{hofstadter}, if the energy levels are plotted versus the flux, see the blue dots in Fig. \ref{fig3} for $N=100$ spins. Even this moderate system size clearly reveals the fractal nature of the spectrum.

In contrast, the energy spectrum for $S_z=N-4$, that is for two spin flips, shown by the red dots in Fig. \ref{fig3}, exhibits significantly less gaps, and covers almost entirely the gapped regions of the butterfly. It is noteworthy that, due to interactions, the energies at $S_z=N-4$ are not simply the sum of the energies at $S_z=N-2$.
For a measurement of the butterfly spectrum, one has to distinguish levels with different magnetization. By a sufficiently strong energy shift via the transverse field $h$, it is possible to energetically separate spectra with different $S_z$. Since the width of the spectrum is of the order of $J$, full separation requires $|h| \gg |J|$. 

\subsection{Topological flat bands}
The appearance of a butterfly spectrum is intimately related to quantum Hall physics and the occurrence of edge states. To visualize edge states in the single-particle spectrum ($S_z=N-2$), we consider the energy levels' dependence on the constant NNN phase $\theta$ for open and periodic boundary conditions, see Fig. \ref{fig2}(a-d). The gaps found for periodic boundary conditions (a+c), are filled by edge states, if the boundary is open (b+d). As seen in Fig. \ref{fig2}(e), the edge states in our 1D system localize at the ends of the spin chain.

Another measure of topological order, routinely applied to two-dimensional systems, are Chern numbers, that is, the winding numbers of
the energy bands subjected to periodic boundary conditions \cite{TKNN}. These numbers are robust in the sense that only a
perturbation which closes the energy gap can modify their integer value. It is possible to adapt the definition of Chern numbers to one-dimensional systems if two parameters are available on which the Hamiltonian periodically depends. In the single-particle configuration ($S_z=N-2$), one parameter is naturally given by the wavevector $k$ of the bands. As a second parameter,  we can take the phase $\theta$. The dependence of the eigenstates $\ket{u(k,\theta)}$ on these parameters is measured by the Berry connection, a two-component vector field defined as $A_{\mu}(k,\theta)= \bra{u(k,\theta)} \partial_\mu \ket{u(k,\theta)}$, where $\mu=\{k,\theta\}$. The Chern number is then obtained as the integral of the Berry curvature over the full parameter space. For those bands which are not separated by a gap, we have to consider the non-Abelian Berry connection, ${\cal A}^{rs}_{\mu}(k,\theta)= \bra{u_r(k,\theta)} \partial_\mu \ket{u_s(k,\theta)}$, where each component is a $N_{\rm deg}\times N_{\rm deg}$ matrix, obtained from
the $N_{\rm deg}$ quasidegenerate bands $\ket{u_r(k,\theta)}$, $r=1,\dots,N_{\rm deg}$. For calculating the Chern numbers, we
follow the method established in Ref. \cite{chern} using a discrete parameter
space. As an example, we have listed the results for $p=1$ and
$q=3,4,5$ in Table \ref{CNtable}.

\begin{table}[t]
\begin{tabular}{c|c}
$q$  & Chern numbers  \\ \hline
 3  & -1, -1, 2, 2, -1, -1  \\
 4 & -1, -1, -1, 6, -1, -1, -1 \\
 5 & -1, -1, -1, -1, 4, 4, -1, -1, -1, -1\\
\end{tabular}
\caption{\label{CNtable}  Chern numbers of gapped manifolds at different
$q$.}
\end{table}

\subsection{Interacting Chern insulator behavior}
Our analysis so far has demonstrated that the single-particle bands, i.e. for
$S_z=N-2$, are topological. However, it is not obvious how these bands are filled if $S_z$ is reduced, as our model is bosonic and becomes strongly interacting in the presence of more than one spin flip. 

\begin{figure}[t]
 \centering
 \includegraphics[width=0.47\textwidth, angle=0]{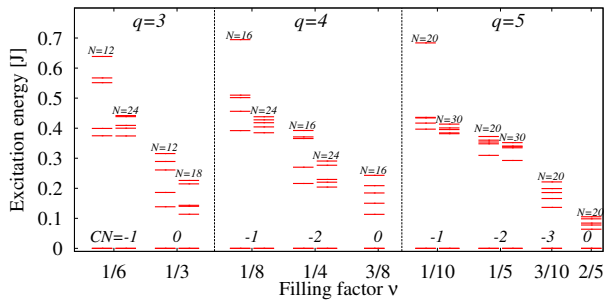}
 \caption{\label{fig4} (Color online)
 Lowest excitation energies for the XY spin chain with $N$ spins at
 different fluxes $\Phi=2\pi/(2q)$, and at different filling factors $\nu=n/(2q)$. The numbers
 above the ground state denote the many-body Chern number (CN).}
 \end{figure}

By exact diagonalization, we have studied the many-body scenario for magnetic fluxes $\Phi=1/(2q)$ at different $S_z$. The most remarkable results are summarized in Fig. \ref{fig4}:
In polarization sectors $S_z/N=1-2\nu$, with $\nu=n/(2q)$, we find gapped ground states for sufficiently small integer values of $n$. These $S_z$ are precisely the configurations which would allow for filling $n$ single-particle bands, and $\nu$ takes the role of a filling factor.

We have checked the topological properties of these states by calculating their Chern number. In the definition of many-body Chern numbers, the wavevector $k$ is replaced by the twist angle of twisted boundary conditions \cite{niu}. Those ground states for which the gap appears to be robust against increasing the system size are found to have non-zero Chern numbers. 
Remarkably, these non-zero Chern numbers correspond to those obtained by filling $n$ single particle bands of Table \ref{CNtable} (compare Table \ref{CNtable} and Fig. \ref{fig4}).
With this, our system provides an example of a strongly interacting Bose system which is topologically equivalent to a non-interacting Fermi system, similar to interacting integer quantum Hall states of two-component Bose gases \cite{senthil2013,furukawa2013,wu2013,davidPRB}.

By spin inversion symmetry our analysis equally holds for the filling factors $\nu=1-n/(2q)$, where $(2q-n)$ bands are filled. On the other hand, no topological order
survives close to half-filling, that is near $S_z=0$: As seen from Fig. \ref{fig4}, the ground state gap becomes smaller near half-filling, and may vanish for large systems. This finding might in principle be due to topological quasi-degeneracies in fractional Chern insulators, but we can exclude this option as the corresponding Chern numbers are zero. For the Hofstadter ladder, a Mott insulating phase is exhibited near $\nu=1/2$ \cite{mariepiraud}.

\section{Experimental implementation}
The realization of long-range spin-spin interactions with trapped ions has been proposed in Ref. \cite{porras04}, and has nowadays become a highly developed experimental routine \cite{schaetz08,monroe10,bollinger2012,blatt14,monroe14,blattross}. A promising new platform are atoms coupled to nanophotonic systems \cite{darrick,kimble14}, with the perspective of better scalability than trapped ions. On both platforms, spin chains with long-range interactions described by the Hamiltonian $H$ of Eq. (\ref{XY}) can be implemented. Naturally, though, the couplings $J_{ij}$ are real-valued, and characterized by either exponential or algebraic decay law. In the following, we describe a driving protocol which modifies the couplings in the desired way. It allows for equipping the system with artificial magnetic fluxes by rendering $J_{ij}$ complex-valued, and we can individually control the interaction strengthes.

The central idea of periodic driving is based on adding fast local potentials to a system. For our purposes, we may simply take
piecewise constant, local energy offsets which can be implemented by applying laser fields that give rise to AC stark shifts \cite{gerritsma}. 
Remarkably, the whole set of information about complex phases and strengthes in the effective couplings can be encoded simply in the times at which the energy offsets change. Our scheme goes beyond those used for atoms in optical lattices, where only nearest-neighbor couplings are relevant.
A precise description of the driving protocol will be given in Appendix B. Here, let us only explain the general idea of our shaking scheme.

Quite generally, the time-dependent Hamiltonian has the form 
\begin{align}
 H(t) = H + \sum_{i} \hbar v_{i}(t)\sigma^{z}_{i}.
\end{align}
With the frequencies $v_i(t)=v_i(t+T)$ varying periodically in time, it is natural to transform the Hamiltonian to the Floquet bases via a gauge transformation $H=U^{\dag}\bar{H} U-iU^{\dag}\dot{U}$ using $U=e^{-i\sum_{i}\chi_{i}(t)\sigma_{i}^{z}}$ with $\chi_{i}(t)=\int_0^t dt' v_i(t')$.
This transformation leaves the Hamiltonian in the XY model form of Eq. (\ref{XY}), but it modifies the coupling parameters.
Provided the periodicity $T$ of the functions $v_i(t)$ is short compared to the time scale of the bare dynamics of the system, the evolution of the system
is well described using time averaged effective couplings
\cite{Eckardt2005}
\begin{eqnarray}
\label{Jtilde}
J_{ij}^{\rm eff}= \frac{J_{ij}}{T} \int_0^T {\rm d}t
e^{2i[\chi_i(t)-\chi_j(t)]}.
\end{eqnarray}
The crucial idea of our shaking protocol is to choose all frequencies $v_i$ to be an integer multiple of some basis frequecy $\nu_0=\frac{ \pi}{\Delta}$. With this, two spins with offsets $\nu_i  \neq \nu_j$ do not accumulate any contribution to the effective coupling $J_{ij}^{\rm eff}$ over a time period which is a multiple of the basic time window $\Delta$, since $\int_t^{t+l\Delta}e^{i 2(\nu_i-\nu_j)t'}{\rm d}t'=0$ for  $l\in \mathbb{N}$. On the other hand, if $\nu_i = \nu_j$ during a time interval $l\Delta$, the strength of the effective coupling, $|J_{ij}^{\rm eff}|$, can be adjusted. Finally, non-trivial phases $\varphi_{ij}=2(\nu_i-\nu_j)\tau_{ij}$ are obtained for $\nu_i \neq \nu_j$ during intervals with a duration $\tau_{ij}$ not being a multiple of $\Delta$. The whole sequence of operations has the duration $T$, and is repeated periodically. These simple basic ingredients are sufficient to implement our model. 

We have worked out a precise protocol, presented in the Appendix B, which allows to engineer our model for arbitrary fluxes in arbirtarily large systems. We use the described scheme to equalize NN and NNN interaction strengthes, and to control the complex phases. We neglect interactions beyond NNN, assuming exponentially fast decaying interactions which are available in atomic systems coupled to nanofibers, or cubically decaying interactions which are available in trapped ions.

Our scheme requires control of energy offsets $v_i$ for individual spins. Typical distances between spins in current experiments with trapped ions are of the order of microns \cite{schaetz08,monroe10,blattross}. Atomic systems coupled to nanophotonic crystals may achieve equally large distances between the spins. Therefore, individual addressing of the spins is no major problem in both platforms. An important experimental requirement concerns the different energy scales which need to be well separated: For the validity of the time averaged description, we require $T^{-1}\gg J_{ij},h$ and the effective coupling rates need to be large compared to the coherence time, $J_{ij}^{\rm eff} \gg t_{\text{dec}}^{-1}$. Also, by demanding $h\gg J$ we shall guarantee a separation of the spectra for different magnetisations $S_z$. In recent trapped ion experiments with tunable interaction range \cite{monroe14,blatt14}, couplings of the order $\sim 100$Hz have been engineered, with coherence times larger than 30ms. Also, local magnetic fields of the order of several kHz were applied. This demonstrates that separation of energy scales is possible in trapped ions. In the Appendix C, we discuss a possible implementation using atoms coupled to nanophotonic crystals~\cite{darrick,kimble14}. There we will also give estimates regarding several constraints which depend on the specific system, for example adiabaticity conditions, if the spin-spin interaction is mediated via photons. 

A possible measurement which can be applied to such spin chain is detection of the energy spectrum.  In this context it is important to notice the presence of $J_{ii}$ terms. Although, for a fixed $S_z$, these terms reduce to a constant, its value still depends on the magnetic flux. Accordingly, the butterfly spectrum appears distorted, but this effect can be removed by purging the spectrum. Edge states could serve as a direct detection of the topological order: Here one needs to measure the energy levels as a function of the tunable parameter $\theta$. For the many-body scenario, detection of the energy gap would already be a strong hint for topological order. For sufficiently small systems, Chern numbers could be reconstructed through full state tomography \cite{roos04,haffner05}.

\section{Summary and Outlook}

In summary, we have studied a mapping between an XY spin chain and particles hopping on a 1D lattice. 
Depending on the spin polarization, the spin chain represents single-particle physics, or a strongly interacting bosonic system.
We have shown that a chain with long-range interactions can be mapped onto a higher-dimensional geometric structure, e.g., a triangular ladder in the case of nearest-neighbor and next-to-nearest-neighbor interactions. Then, the presence of complex-valued interactions can give rise to loops with non-vanishing magnetic flux even in one dimension, and the system potentially exhibits similar physics as the Hofstadter model for a charged particle in a plane with perpendicular magnetic field. In particular, we have found for the single-particle configuration a fractal energy spectrum, edge states, and a band structure with non-zero Chern numbers. For certain many-body configurations, the ground state remains topological with non-zero Chern numbers. It is topologically equivalent to fermionic filling of the single-particle bands, although the system is strongly interacting and bosonic. 

The spin chain model can be engineered in systems of trapped ions or atoms in waveguides. We provide a simple but powerful driving protocol to engineer the artificial magnetic field, and to control the strengthes of the couplings. Due to the high degree of controllability in these systems detection of the fractal energy spectrum or of edge states is feasible.

Our analysis is related also to spin chains with Dzyaloshinksii-Moriya interactions \cite{dzyaloshinskii,moriya,SU3DM}, possibly with long-range character \cite{long-range-DM}. It is also closely connected to studies of ladder systems incorporating the Hofstadter model \cite{paredes-ladder,bloch-ladder,grusdt14,mariepiraud}, and to the Hofstadter model with long-range hopping \cite{hatsugai90}.

We have focused on the case with nearest-neighbor and next-nearest neighbor interactions. In view of the tunability of the interactions, a systematic study of the influence of longer-range interactions could be a relevant extension of our work. Another interesting aspect is the extension from an XY model to a Heisenberg model: The $S_z-S_z$ interaction can provide an additional
repulsion between the spin flips, which could give rise to fractional Chern insulator behavior.

\appendix

\section{Harper-like equation}
If all but one spins are polarized, the Hamiltonian (\ref{XY}) can be solved by Fourier transformation. Assuming  non-zero NN and NNN couplings and magnetic fluxes $\Phi=\pi p/(2q)$, as shown in Fig. \ref{fig1}, the wave vectors should be restricted  to a magnetic Brillouin zone, $k \in [-\pi/2q,\pi/2q]$. Distinguishing between odd ($\ell=1$) and even sites ($\ell=2$), we introduce Fourier-transformed spin-flip operators $\sigma_{k\nu\ell}^\pm$, with $\nu$ the band index and $k$ the wave vector. With this, we obtain a Harper-like Hamiltonian \cite{harper} given by 
\begin{align}
 H=& -J \sum_{k,\nu} \Big[ e^{i(k+\pi\nu\frac{p}{q})}
\left(
\sigma_{k\nu1}^+,
\sigma_{k\nu2}^+
\right)
\left(
\begin{matrix}
 e^{i\theta} & 1 \\
 0 & e^{i\theta}
\end{matrix}
\right)
\left(
\begin{matrix}
 \sigma_{k\nu1}^- \\
 \sigma_{k\nu2}^-
\end{matrix}
\right)
+ \nonumber \\ &
 \left(
\sigma_{k\nu1}^+,
\sigma_{k\nu2}^+
\right)
\left(
\begin{matrix}
 0 & 0 \\
 1 & 0
\end{matrix}
\right)
\left(
\begin{matrix}
 \sigma_{k(\nu+1)1}^- \\
 \sigma_{k(\nu+1)2}^-
\end{matrix}
\right) \Big] +{\rm H.c.},
\end{align}
Since $\sigma_{k\nu\ell}^\pm= \sigma_{k(\nu+2q)\ell}^\pm$, the Hamiltonian can be written as a $2q \times 2q$ matrix. Diagonalizing this matrix yields the energy bands.

\vspace{1cm}
\section{Driving protocol}
Here we propose a specific shaking protocol which, for arbitrarily large chains, allows for engineering the couplings $J_{ij}^{\rm eff}$ according to our needs. The only ingredient to our protocol are local energy offsets $v_j(t)$ which alternatingly take two values, $\nu_j \neq 0 $ or zero. The frequencies $\nu_j$ are multiples of a base frequency $\nu_0$, chosen such that during an elementary time window $\Delta$ no contribution is made to $J_{ij}^{\rm eff}$ for $\nu_i \neq \nu_j$. In our protocol, we use energy offsets $\nu_j=\nu_0,\ \!2\nu_0,\ \!3\nu_0,\ \!4\nu_0,\nu_0,\ \!2\nu_0$, ... along the chain, that is, $\nu_j=\left(j\text{mod}4+4\delta_{j\text{mod}4,0}\right)\cdot\nu_0$.
All information about complex phases and strengthes of the effective couplings $J^{\rm eff}_{ij}$ can then be encoded in the times at which the energy offsets of the spins $i$ and $j$ are set to zero, as explained below.

\begin{figure}[t]
\centering
\includegraphics[width=0.49\textwidth, angle=0]{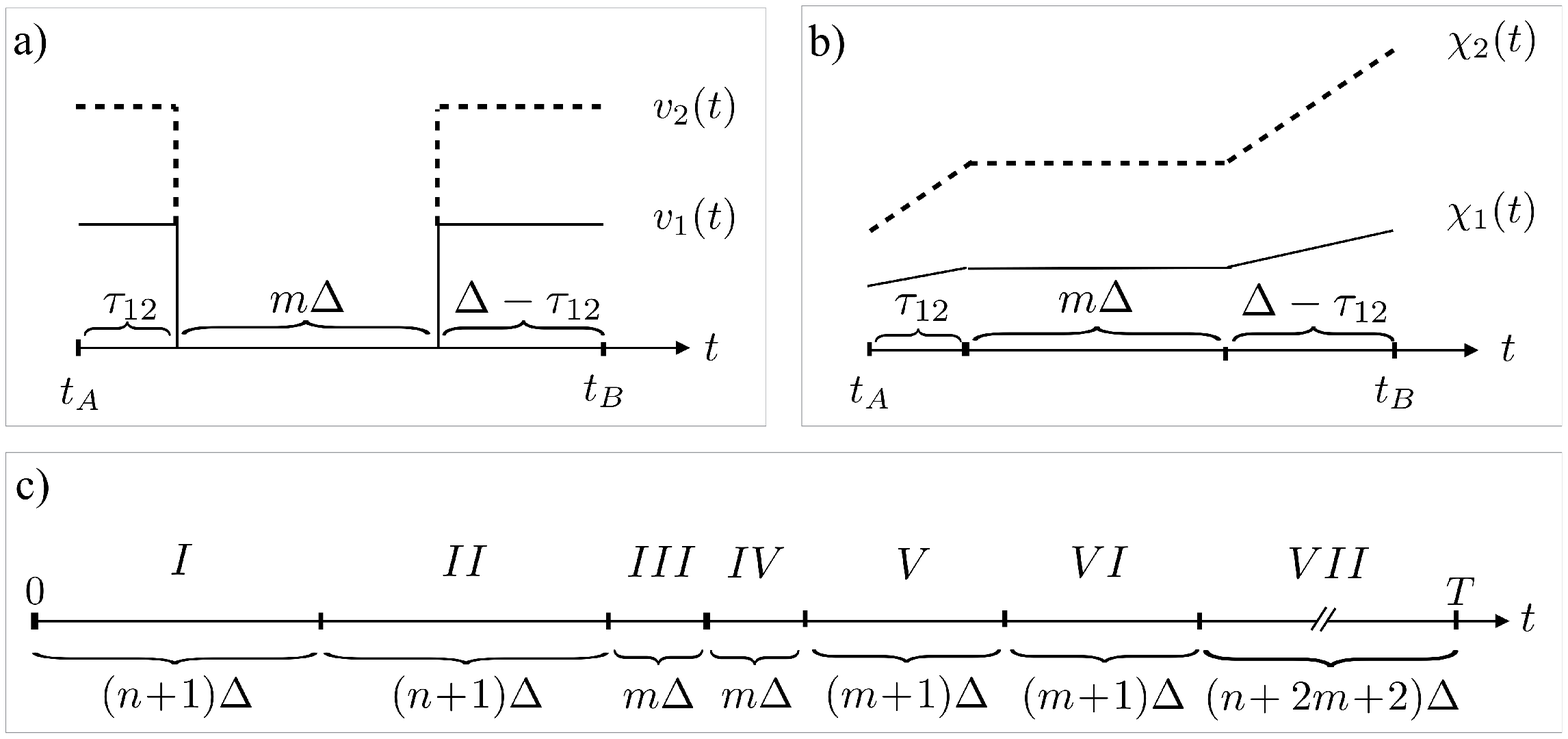}
\caption{\label{fig5}
(a) Time-dependence of the local energy offsets $v_i$
on sites $1$ and $2$ during an elementary sequence. (b) Integrated energy offsets $\chi_i$
which give rise to complex hopping parameters, see Eq. (\ref{Jtilde}). (c) The
whole shaking period $T$ and its division into seven parts is depicted. As explained in the text, the elementary sequences I-II control the NNN couplings. Sequences III--VI control the NN couplings, and sequence VII is used for averaging all energy offsets to zero.}
\end{figure}

As illustrated in Fig.~\ref{fig5}c, the shaking cycle which is repeated periodically consists of seven time intervals. The time intervals I - VI are used for defining the effective couplings $J_{ij}$.  The last time interval (VII) ensures that the time-average of the shaking function is zero, $\frac{1}{T}\int_0^Tdtv_{i}(t)=0$. To this end, the energy-offset of each spin takes the constant negative value $-\nu_i$ during this time interval. The explicit definitions of the shaking functions $v_i(t)$ in each interval are given in Table II. 

In each of the intervals I--VI, different spin pairs $i$ and $j$ are addressed via the simple, elementary sequence shown in Fig. \ref{fig5}ab: Time interval I is used to tailor the interactions that correspond to the upper horizontal links in Fig.~\ref{fig1}, that is, $J_{i,i+2}$ with $i$ odd. Accordingly, only spins with odd indices are addressed during this interval. Time interval II is used to tailor the interactions that correspond to the lower horizontal links in Fig.~\ref{fig1}, that is, $J_{i,i+2}$ with $i$ even.  Therefore, only spins with even indices are addressed. The following four sections (III-VI) serve for defining the NN couplings, where we distinguish between the real-valued couplings $J_{2i,2i+1}$ (red links in Fig.~\ref{fig1}) addressed in blocks III (for even $i$) and IV (for odd $i$), and the complex-valued couplings $J_{2i-1,2i}$ (blue links in Fig.~\ref{fig1}) addressed in blocks V (for odd $i$) and VI (for even $i$). In all cases, the distinction between even and odd $i$ is crucial for avoiding undesired contributions, as it guarantees that any pair $i$ and $j$ addressed simultaneously during one interval either corresponds to a coupling $J_{ij}$ which is supposed to be adjusted, or has a distance $|i-j|>2$ which allows to neglect the corresponding coupling $J_{ij}$ due to the fast decay of the bare couplings. 

\begin{table}[t]
\begin{tabular}{c|c}
$v_j(t)$  & time intervals (in units of $\Delta$) \\ \hline
 0  & $\left[\!(j\!+\!1)\text{mod}2\!\cdot \!(n\!+\!1)\!+\!\frac{\theta}{4\nu_0\Delta},
\left\{\!(j\!+\!1)\text{mod}2\!+\!1\right\}\!(n\!+\!1)\!+\!\frac{\theta}{4\nu_0\Delta}\!\right]$ \\
0 & $[ 2(n+1)+m\cdot \Theta\left(j\text{mod}4-1.5\right),$ \\
& $ 2(n+1)+m\cdot \Theta\left(j\text{mod}4-1.5\right)+m ]$ \\
0 & $[ 2(n+m+1)+(m+1)\cdot \Theta\left((j-1)\text{mod}4-1.5\right)+\frac{t_{j}}{\Delta},$ \\ &
$ 2n+3m+3+(m+1)\cdot \Theta\left((j-1)\text{mod}4-1.5\right)+\frac{t_{j}}{\Delta}]$ \\
 $-\nu_j$ & $\left[2(n+2m+2),\
3(n+2m+2)\right]$ \\
 $\nu_j$ & for all other times\\
\end{tabular}
\caption{\label{protocol}  Shaking functions $v_j(t)$ for realizing the
interactions shown in Fig.~\ref{fig1}.  $\Theta(t)$ is the
Heaviside function, $\nu_j=\left[ j\text{mod}4+4\delta_{j\text{mod}4,0} \right]\nu_0$, and
$t_j=j\text{mod}2\cdot \frac{\varphi_{j,j+1}}{2v_0}+(j+1)\text{mod}2\cdot\frac{\varphi_{j-1,j}}{2v_0}$
for arbitrary $\varphi_{ij} \in [0,2\pi]$.
Line 1 in the table corresponds to time intervals I/II in Fig. \ref{fig5}(c), line 2 corresponds to time intervals III/IV, line 3 corresponds to time intervals V/VI, line 4 corresponds to time interval VII.
}
\end{table}

In each time interval I-VI,  the elementary sequence shown in Fig. \ref{fig5}ab is performed. We will now show that simply the timing of the energy drops determines the effective phases and strengthes. Without loss of generality, we assume that the elementary sequence occurs within a time interval $ [t_A, t_B] $ with $t_A\text{mod}\Delta\! = 0$  and $t_B\!=\!t_A\!+\!(n\!+\!1)\Delta$ with $n\in \mathbb{N}$. For this time interval, the shaking functions of two spins $i$ and $j$ are given by $v_i(t) = \nu_i f(t)$ and $v_j(t)=\nu_j f(t)$, with $f(t)=1-\Theta(t-[t_A+\tau_{ij}])+\Theta(t-[t_A+\tau_{ij}+n\Delta])$, where $\Theta(t)$ is the Heaviside step function. In the function $f(t)$, the parameter $\tau_{ij}$ determines when the energy offset drops to zero, while the (integer) parameter $n$ determines its duration at zero. Let us now parametrize $\tau_{ij}$ by  $\tau_{ij}=\frac{\varphi_{ij}}{2(\nu_i-\nu_j)}$ (assuming $\nu_i \neq \nu_j$). With this, we obtain
\begin{align}
 \int_{t_A}^{t_B} e^{i2[\chi_i(t')-\chi_j(t')]}dt=e^{i\varphi_{ij}} n\Delta.
\end{align}
From this expression we find that the complex phase $\varphi_{ij}$ of the coupling is controlled by the timing $\tau_{ij}$, and the duration at zero, $n\Delta$, determines the coupling strength.

As shown in Fig.~\ref{fig5}c, the time intervals for tuning different couplings have different durations. We shall note that in intervals I,II,V, and VI, one time unit $\Delta$ is spent for adjusting the complex phase. This leaves $n$ time units in intervals I and II and $m$ time units in intervals III-VI available for tuning the interaction strength. By setting the energy offsets equal to zero during time windows of length $n$ and $m$ respectively, we can achieve a ratio between effective NN couplings $J_1^{\rm eff}$ and effective NNN couplings $J_2^{\rm eff}$ given by $\frac{J_2^{\rm eff}}{J_1^{\rm eff}}=\frac{n}{m}\cdot\frac{J_2}{J_1}$.


\vspace{1cm}
\section{Implementation in nanophotonic systems}
Cold atoms trapped near one-dimensional photonic crystals as described in~\cite{darrick,kimble14} are a very promising platform for realizing strong and tuneable long-range interactions for a large number of atoms. The spin states $|g\rangle$ and $|s\rangle$ can be encoded in ground states of atoms with a lambda configuration, where the excited state is adiabatically eliminated. The photonic modes in the crystal mediate an effective atom-atom interaction of the type given in Eq.~(\ref{XY}), with exponentially decaying coupling strength $J_{ij}=J \ \! e^{-\frac{2| r_i-r_j|}{L}}$, where $r_{i/j}$ are the positions of the atoms with indices $i$, $j$, and $L$ is the characteristic length of the interaction. The range of the coupling $L$  is tunable through a variation of the system parameters (see~\cite{darrick} for details).
The ratio of the parameters $n$ and $m$ in Fig.~\ref{fig5}c takes here the value $\frac{n}{m}=e^{\frac{2a}{L}}$, where $a$ is the atomic spacing. 
Undesired terms with $|i-j|>2$ are suppressed by a factor $e^{-\frac{4a}{L}}$ or more. An even stronger suppression can be achieved by introducing more time intervals.

\paragraph{Losses.}
The system is subject to losses which limit the coherence time. The two main loss channels are the spontaneous emission of the atoms into free space and the loss of photons due to imperfections of the photonic crystal. As described in~\cite{darrick}, the atoms coupling to light modes in the crystal can be treated in analogy to a cavity QED system with cavity length $L$. For optimised detuning of the classical driving field, the ratio of the coherent interaction rate $J$ to the rate at which the system looses excitations $\Gamma$ is determined by the cooperativity $C$, 
\begin{eqnarray}\label{Eq:MinimalLosses}
\frac{\bar{J}_0}{\Gamma}=\sqrt{C}.
\end{eqnarray}
The cooperativity depends on the characteristic length $L$, $C=\frac{L}{\lambda}C_{\lambda}$, where $\lambda$ is the resonant wavelength of the atomic transition ($\lambda=2\pi c/\omega_{ge}$.
Depending on the quality factor $Q$ of the photonics crystal, $C_{\lambda}$ can take values up to $C_{\lambda}=10^4$ (for $Q=10^{6}$) or $C_{\lambda}=10^5$ (for $Q=10^{7}$).
For example, a configuration with a distance $a=\frac{3}{2}\lambda$ between the atoms and  $L=\frac{3}{\ln(3)}  \lambda$ can be implemented with $n=3$, $m=1$. This yields an effective cooperativity of $C_{\text{eff}}=80$ for $Q=10^7$, where we used $\sqrt{C_{\text{eff}}}=J_0/\Gamma=\sqrt{C}\frac{1}{3(n+2m+2)}$. Hence it should be possible to implement predominantly coherent effective interactions in a system consisting of $\mathcal{O}(10^2)$ atoms.\\
\paragraph{Required energy scales.}
In the following, we discuss briefly the required parameter regimes. The time averaged description is valid for $\frac{1}{T}\gg  \left(J,\ h\right)$. Therefore the condition
\begin{eqnarray*}
v_0\gg 3 \pi(n+2m+2)\left(J,\ h\right)
\end{eqnarray*}
must be fulfilled. Moreover, the shaking procedure must be compatible with the conditions under which the interaction Hamiltonian given in Eq.~(\ref{XY}) is valid. As described in~\cite{darrick}, this Hamiltonian is obtained by eliminating the light field in the photonics crystal and the excited state $|e\rangle$. In order to prevent real excitations of the photon field mediating the interactions and in order to avoid the population of the excited atomic states, the maximum energy offset in the shaking protocol has to be small compared to the detuning $\delta_L$ of the applied classical laser field with respect to the atomic transition, $\delta_L\gg 4\nu_0$. Finally, the effective couplings have to be large compared to the decoherence rate $J_{ij}^{\rm eff}\gg t^{-1}_{\text{dec}}$. For the following estimate, we assume the decoherence time to be limited by the loss rate $\Gamma$ discussed above. In summary, we require
\begin{eqnarray}
& \frac{\delta_L}{4}\gg v_0\gg 3 \pi(n+2m+2) \left(J, h\right)\gg \frac{9 \pi(n+2m+2)^2}{t_{\text{dec}}}
\nonumber \\
& \sim \frac{9 \pi(n+2m+2)^2}{\sqrt{C}}J.
\label{Eq:Hierarchies}
\end{eqnarray}
The bare coupling parameter $J$ is given by $J=\frac{|\Omega|^2\bar{g}^2_c}{2 \Delta_L \delta_L^2}$, where $\bar{g}_c$ is the light-atom coupling constant. The parameter  and $\Delta_L$ is a detuning (see \cite{darrick} for details) which has to be optimised in order to guarantee a minimal loss-rate as stated in Eq.~(\ref{Eq:MinimalLosses}) yielding $J= \frac{|\Omega|^2}{\delta_L^2}\sqrt{C} \ \!\gamma$. This expression is used as an estimate, but the actual optimisation is more complicated since $\Delta_L$ plays also a role for tuning the characteristic length $L$. One finds
\begin{eqnarray*}
J \approx \frac{|\Omega|^2}{\delta_L^2}\sqrt{C} \ \!\gamma,
\end{eqnarray*}
where $\Omega$ is the Rabi frequency of the applied laser field and $\gamma$ is the atomic line width (for Cesium atoms $\gamma/(2\pi)\sim 5$MHz).
Using $a=\frac{3}{2}\lambda$ and $L=\frac{3}{\ln(3)}  \lambda$, as above ($n=3$, $m=1$), we find that the conditions in Eq.~(\ref{Eq:Hierarchies}) can be fulfilled for a sufficiently large detuning $\delta_L$.
This estimate suggests that the protocol could be realized in principle in its basic form described above. By optimizing the scheme, an enhanced performance and increased robustness can be achieved.

\section*{Acknowledgments} We wish to thank Emil Bergholtz, Darrick
Chang, James Douglas, Philipp Hauke, and Christian Roos for fruitful
discussions and correspondences. We acknowledge support from EU grants OSYRIS (ERC-2013-AdG Grant No. 339106), SIQS (FP7-ICT-2011-9 No. 600645), QUIC (H2020-FETPROACT-2014 No. 641122), EQuaM (FP7/2007-2013 Grant No. 323714), Spanish Ministry grant FOQUS (FIS2013-46768-P), and Fundació Cellex. R.W.C. acknowledges a Mobility Plus fellowship from the Polish Ministry of Science and Higher Education, and the Polish National Science Center Grant No. DEC-2011/03/B/ST2/01903.

\medskip


\end{document}